%% file: hybrid_compilation.tex
\documentclass[%
superscriptaddress,
 amsmath,amssymb,
 aps,
pra,
]{revtex4-2}

\usepackage{hyperref}
\hypersetup{colorlinks=true}

\usepackage{graphicx}
\usepackage{dcolumn}
\usepackage{bm}

\usepackage{cleveref}
\crefname{figure}{Fig.}{Figs.}
\crefname{listing}{program code}{program codes}
\Crefname{listing}{Program Code}{Program Codes}
\Crefformat{appendix}{#2Appendix~#1#3}

\usepackage[frozencache=true,cachedir=.,newfloat]{minted} 
\SetupFloatingEnvironment{listing}{name=Program Code}
\SetupFloatingEnvironment{listing}{listname=List of Program Code}
\SetupFloatingEnvironment{listing}{placement=hbpt}

\usepackage{xcolor}
\definecolor{bg}{rgb}{0.95,0.95,0.95}

\usepackage{textcomp} 

\newcommand{\Nq}{N_\mathrm{q}}

\usepackage{physics}

\begin{document}

\title{An LLVM-based C++ Compiler Toolchain for Variational Hybrid Quantum-Classical Algorithms and Quantum Accelerators}

\author{Pradnya Khalate}
\thanks{Both authors contributed equally to this work}
\email{xin-chuan.wu@intel.com}
\affiliation{Intel Labs, Intel Corporation, 2111 NE 25th Ave, Hillsboro, OR 97124}

\author{Xin-Chuan Wu}
\thanks{Both authors contributed equally to this work}
\email{xin-chuan.wu@intel.com}
\affiliation{Intel Labs, Intel Corporation, 2200 Mission College Blvd, Santa Clara, CA 95054}

\author{Shavindra Premaratne}
\affiliation{Intel Labs, Intel Corporation, 2111 NE 25th Ave, Hillsboro, OR 97124}
\author{Justin Hogaboam}
\thanks{Present Affiliation: Microsoft Corporation, Redmond, WA, USA}
\author{Adam Holmes}
\thanks{Present Affiliation: HRL Laboratories, Malibu, CA, USA}
\author{Albert Schmitz}
\affiliation{Intel Labs, Intel Corporation, 2111 NE 25th Ave, Hillsboro, OR 97124}

\author{Gian Giacomo Guerreschi}
\affiliation{Intel Labs, Intel Corporation, 2200 Mission College Blvd, Santa Clara, CA 95054}

\author{Xiang Zou}
\author{A. Y. Matsuura}
\affiliation{Intel Labs, Intel Corporation, 2111 NE 25th Ave, Hillsboro, OR 97124}

\date{\today}

\begin{abstract}
Variational algorithms are a representative class of quantum computing workloads that combine quantum and classical computing. This paper presents an LLVM-based C++ compiler toolchain to efficiently execute variational hybrid quantum-classical algorithms on a computational system in which the quantum device acts as an accelerator. We introduce a set of extensions to the C++ language for programming these algorithms.  We define a novel Executable and Linking Format (ELF) for Quantum and create a quantum device compiler component in the LLVM framework to compile the quantum part of the C++ source and reuse the host compiler in the LLVM framework to compile the classical computing part of the C++ source.  A variational algorithm runs a quantum circuit repeatedly, each time with different gate parameters.  We add to the quantum runtime the capability to execute dynamically a quantum circuit with different parameters.  Thus, programmers can call quantum routines the same way as classical routines.  With these capabilities, a variational hybrid quantum-classical algorithm can be specified in a single-source code and only needs to be compiled once for all iterations.  The single compilation significantly reduces the execution latency of variational algorithms.  We evaluate the framework's performance by running quantum circuits that prepare Thermofield Double (TFD) states, a quantum-classical variational algorithm. 

\end{abstract}

\maketitle

\section{Introduction\label{sec:introduction}}

\begin{figure}[ht]
\includegraphics[width=0.52\columnwidth]{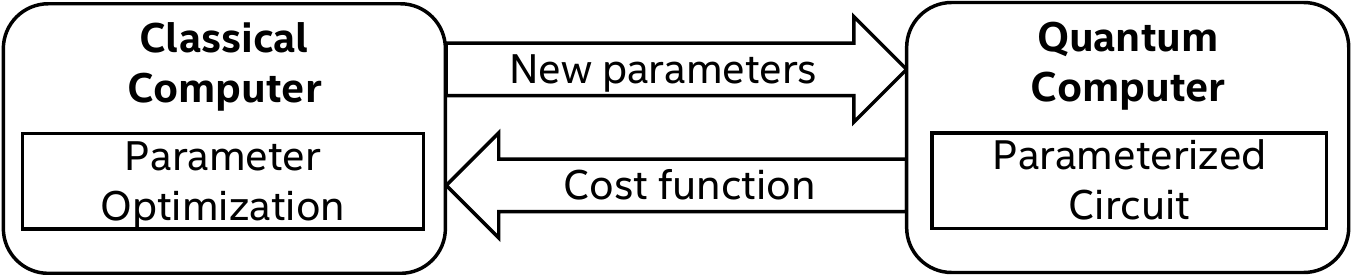}
\caption{Optimization loop in hybrid quantum-classical variational algorithms. The values of the quantum circuit's parameters vary for each iteration, and depend on the results of the previous optimization loop. \label{fig:optimization_loop}}
\end{figure}

Quantum computing has the potential to solve classically intractable problems \citep{Montanaro2016}. Current quantum computers in the Noisy Intermediate-Scale Quantum (NISQ) \citep{Preskill2018} regime are limited by the total number of operations that can be performed reliably, as well as the total number of quantum bits (qubits) available. Hybrid quantum-classical algorithms \citep{Cerezo2021} such as the Variational Quantum Eigensolver (VQE) \citep{Wecker2015} and the Quantum Approximate Optimization Algorithm (QAOA) \citep{Farhi2014} are amongst the most promising near-term methods for addressing applications areas like materials, chemistry, and industrial optimization. These applications leverage classical computers by delegating part of the workload to a CPU (central processing unit) and depend on a tight relationship between classical computers and quantum accelerators during program execution. \Cref{fig:optimization_loop} illustrates a quantum-classical optimization loop to optimize the parameters and evaluate candidate solutions of increasing quality. The parameters in the quantum circuit of each iteration depend on the previous results in the optimization loop. During this optimization loop, the CPU calculates a new set of variational parameters to be used  for the next run of the algorithm, and makes them available to the components responsible for generating the quantum operations. Here we focus on the compiler and runtime techniques that allow programmers to design hybrid quantum-classical variational algorithms in a single C++ source code and generate a binary executable, which performs classical computation on the host CPU and quantum computation on the quantum accelerator.

Several studies related to quantum software platforms and compilers have been carried out. Many software development kits are implemented with Python and C++. The platforms built based on Python enable the capabilities of analyzing and optimizing NISQ applications. For example, IBM Qiskit \citep{cross2018ibm}, Rigetti Forest \citep{rigetti}, and Google Cirq \citep{Cirq} provide their frameworks for compilation and execution of quantum programs. Xanadu's Strawberry Fields \citep{killoran2019strawberry} offers the Python library for photonic quantum computing. TILT \citep{wu2021tilt} compiles quantum circuits for trapped-ion systems. PyZX \citep{kissinger2019pyzx} performs quantum circuit optimization using the ZX-calculus. BQSKit provides a Python library to synthesize quantum circuits \citep{younis2021berkeley, wu2020qgo}. Tequila \citep{Tequila2021} is a Pythonic package used for prototyping many aspects of hybrid algorithms, from ansatz choices to measurement reductions. Staq \citep{amy2020staq} provides a C++ library for the synthesis and compilation of quantum applications. ScaffCC \citep{JavadiAbhari2015} provides a framework for compilation, analysis, and optimization. However, none of these frameworks support a runtime library. A noticeable exception is QCOR \citep{mintz2020qcor, mccaskey2020xacc}, a C++ language extension for hybrid quantum-classical programming which provides runtime library to support the quantum program execution. However, QCOR requires re-generating multiple quantum circuits to support the execution of variational algorithms. When handling scalable quantum applications, the frameworks based on C++ have demonstrated significant performance advantages compared to Python-based frameworks \citep{Litteken2020}. 

In this work, we introduce a C++ framework that is designed for optimization, compilation, and execution of scalable hybrid quantum-classical variational algorithms. Since our framework supports dynamic parameters, a variational algorithm is only compiled once for the execution of all iterations.  Our quantum extensions to the C++ programming language leverages the LLVM infrastructure to compile a quantum program. This work adopts the programming language model originally demonstrated by ScaffCC \citep{JavadiAbhari2015} to the extent of defining custom datatypes, custom intrinsic functions, and utilizing the LLVM pass infrastructure. The primary goals of our implementation are flexibility of expressing hybrid quantum-classical algorithms, and enabling support for handling dynamic parameters with a single compilation. This includes a code generator for our qubit control processor, a quantum runtime library for managing execution, and the definition of an application binary interface (ABI) for executable files.


A compiler is an application that translates a program written in the source language (\emph{source program}) to an equivalent program in the target language (\emph{target program}) such as an executable \citep{Aho2006}. Generating an executable from source code is a multi-phase process that can be broadly categorized into three stages: language parser and analyzer (\emph{front}end); optimizer (\emph{middle}end); and code generator (\emph{back}end). LLVM is a collection of modular and reusable compiler toolchain technologies which uses this three-phase design \citep{LLVM}. It includes Clang -– a compiler for the C family of languages, a target-independent optimizer, an extensible code generator for new targets, and a linker. The frontend parses the input source program written in a high-level programming language, performs lexical, syntactic, and semantic analyses, and on success, generates a lower-level intermediate representation (IR) as the output. The optimizer stage is responsible for improving the execution performance. In LLVM, this is managed by transformation passes which translate the IR into an optimized IR. The code generator is associated with a specific target machine, and performs code generation by mapping the IR to the target instruction set.

\begin{figure}[ht]
\includegraphics[width=0.42\columnwidth]{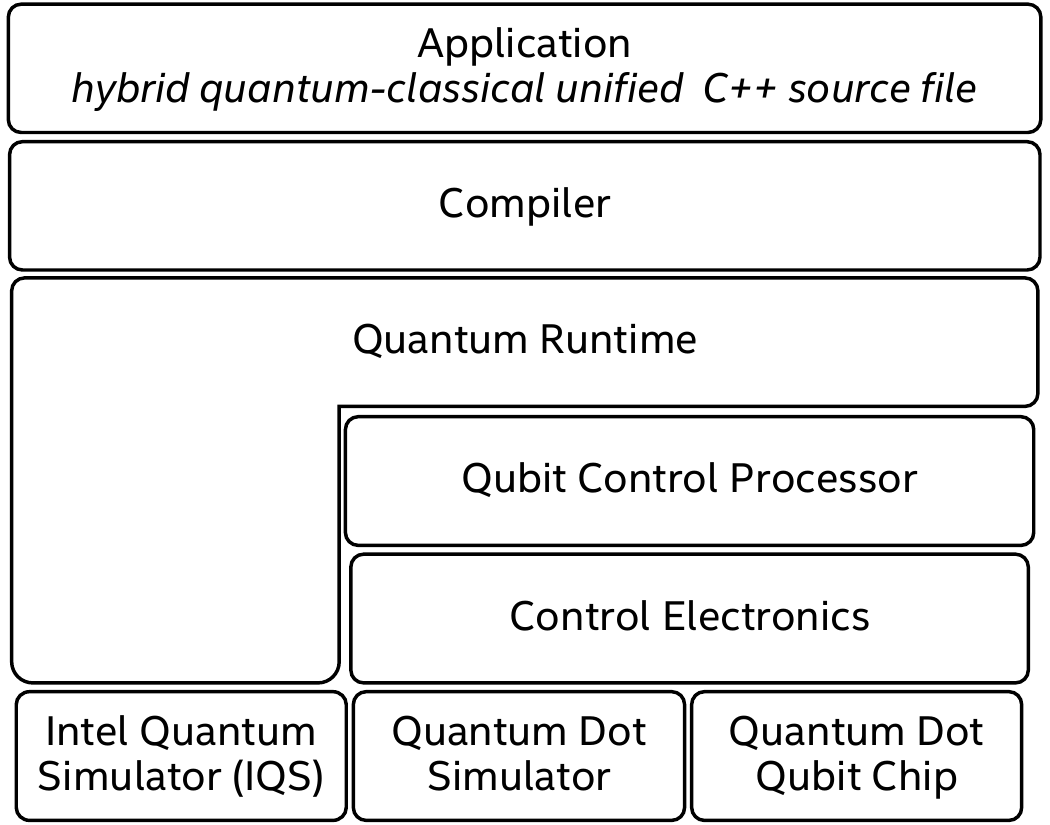}
\caption{Layered architecture of Intel Quantum Computing stack. Adopting a modular design with well-defined interfaces between each layer makes it an extensible system.\label{fig:stack}}
\end{figure}

To enable the power of quantum computation, we propose a full-stack of system software and hardware. \Cref{fig:stack} illustrates the layered architecture of our full-stack quantum computing system which includes an LLVM-based quantum compiler toolchain, quantum runtime, qubit control processor, control electronics, and qubit devices (targets) performing quantum computation, with well-defined interfaces. The modular design allows us to support multiple components in each layer and replace any component easily in the future. The application is a quantum algorithm and any relevant classical logic represented as a C++ program. The compiler translates this unified C++ source file into a binary executable. The quantum runtime provides library calls for managing quantum-classical interaction and communicating with the qubit control processors that manage the execution on the specified qubit target. The target can be an actual quantum processor (\textit{e.g.} a quantum dot qubit chip), or a simulator reproducing  the behavior of quantum processors in software. A quantum circuit simulator like the Intel Quantum Simulator (IQS) \citep{Guerreschi2020}, which is agnostic to the specificity of the physical implementation of the qubits, can directly interact with the quantum runtime interface and execute the quantum circuit. A qubit chip or its equivalent simulator additionally needs the qubit control processor and supporting electronics for generating the necessary control signals. In this paper, we will focus on the top three layers and their implementation.

To summarize, our main contributions are as follows:
\begin{itemize}
\item We extend the LLVM framework to enable the compilation of quantum programs. This allows programmers to leverage existing compiler techniques for quantum program analysis, optimization, and executable code generation.
\item We propose a compilation and execution model for hybrid quantum-classical variational algorithms. With our approach, the hybrid program only needs to be compiled once for the execution of multiple algorithm iterations. This approach reduces the overall latency from $N_{it}\dot(T_{c} + T_{e})$ to $T_{c} + N_{it}T_{e}$, where $N_{it}$ is the number of iterations, and $T_c$ and $T_e$ are the compilation time and execution time, respectively. 
\item We implement the full-stack software platform including C++ quantum extensions, compiler, runtime, and qubit simulator. We demonstrate the feasibility of our approach by running a hybrid quantum-classical algorithm that prepares purified versions of thermal equilibrium states known as Thermofield Double (TFD) states \citep{Wu2019}. This variational TFD algorithm is important to the application area of materials design or the simulation of complex electronic materials.
\end{itemize}

\section{Features of Intel Quantum Compiler Toolchain\label{sec:iqc_overview}}

\subsection{Single Compiler for Both Classical and Quantum Programs}

The Intel quantum compiler toolchain provides a software platform that allows programmers to design, compile, and execute hybrid quantum-classical applications. The ability to intermix quantum and classical code facilitates tighter coupling and ease of programmability. Our design enables users to use any existing C++ libraries for the classical computation on the host CPU. 
This compiler toolchain comprises of two primary compiler components. 
\begin{enumerate}
\item A host compiler which is a standard LLVM-based Clang compiler for the CPU.
\item A device compiler which is targeted for the specifications of our quantum instruction set architecture (QuISA) \citep{Fu2017, Zou2020} for the quantum accelerator.
\end{enumerate}

\begin{figure}[ht]
\includegraphics[width=0.42\columnwidth]{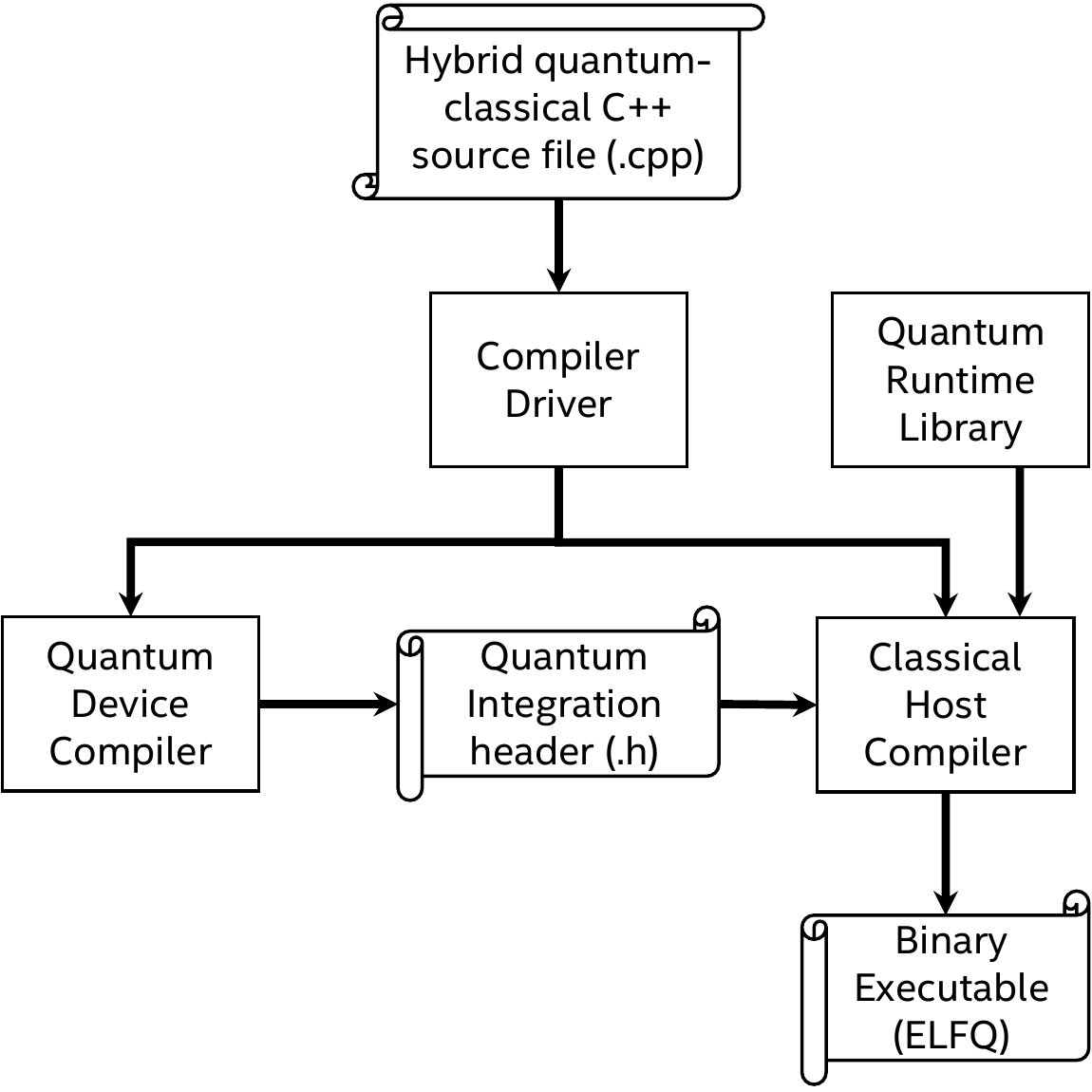}
\caption{Intel quantum compiler toolchain's hybrid compilation workflow. The components of the compiler toolchain are represented with boxes, and the input-output files with scrolls. User input is distinguished with a vertical scroll and compiler-generated files as horizontal scrolls. The application source file is processed by the compiler driver to extract the quantum logic and generate integration header file which is merged back into the classical code in order to produce a unified hybrid executable binary file by linking with the Quantum Runtime (QRT) library.\label{fig:hybrid_flow}}
\end{figure}

\Cref{fig:hybrid_flow} shows the overview of our design to support the compilation of hybrid quantum-classical variational algorithms. The input to the compiler is a single C++ source file which represents the application layer, including classical functions and quantum kernels. The compilation process is orchestrated by the compiler driver which logically splits the process into quantum device compilation flow and classical host compilation flow, and invokes the necessary components in the toolchain to process the input. The compiler frontend extracts the quantum kernels from the unified source code and passes it to the quantum device compiler. The device compiler performs analysis and optimization for the quantum kernels, and generates an integration header file that provides quantum kernel information, such as function names, symbols, function lengths, and macroinstructions for the runtime library. The host compiler receives the quantum integration header and merges back the quantum logic with the classical logic and links with the quantum runtime library. Once all compilation processes are completed, the output binary executable is created in our novel Executable and Linking Format \citep{TIS_ELF} for Quantum (ELFQ). A single ELFQ file can contain all the required information to run a hybrid quantum-classical application.

\subsection{Hybrid Execution}
\begin{figure}[ht]
\includegraphics[width=0.96\columnwidth]{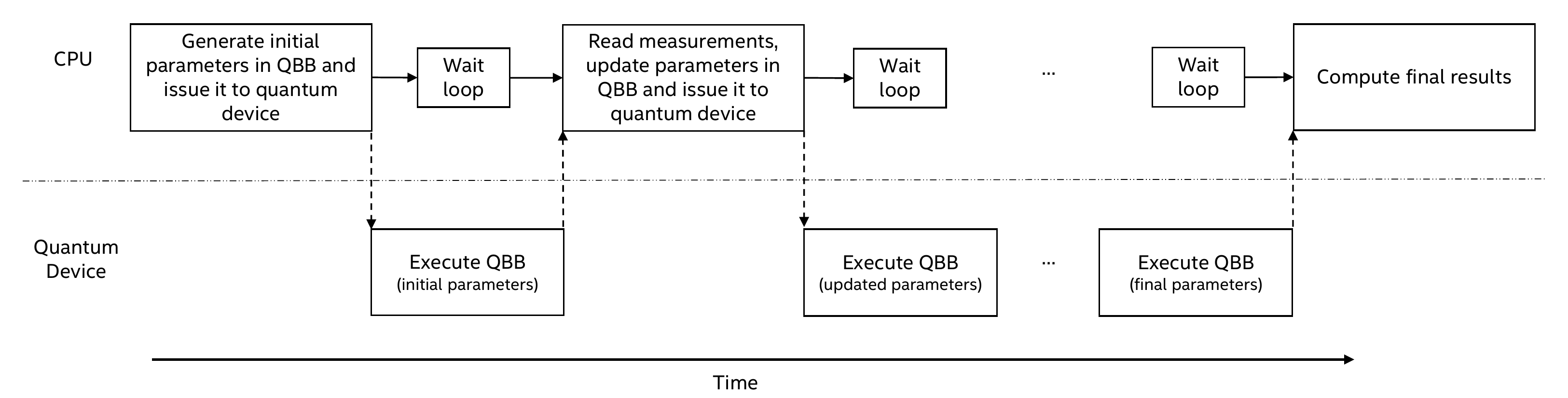}
\caption{Execution flow between the host CPU and the quantum device for running hybrid applications with dynamic parameters.\label{fig:control_flow}}
\end{figure}

In our runtime design for hybrid execution, we introduce a new concept called the \emph{quantum basic block} (QBB) that is a linear sequence of quantum operations. The quantum kernels are compiled into multiple QBBs. \Cref{fig:control_flow} shows our runtime execution  flow. The program is launched by the host CPU with an initial set of values for the dynamic parameters used in the quantum program. Once a quantum kernel is called by the program, the host system uses a blocking call to issue the corresponding QBB to the quantum system and waits for the quantum device. After the QBB execution is complete, the results are passed to the host CPU. The host CPU generates an updated set of values for the dynamic parameters used in the quantum program and issues the next QBB. This process continues until the final results are computed by the host CPU.

It is important to note that our design does not require quantum circuits to be composed by a single QBB. There are several reasons why circuits with multiple QBBs are desirable. First, this may reflect the logic of the quantum circuit; for example separating the preparation of a reference qubit state from its subsequent manipulation. Second, complex quantum programs may require sequencing QBBs in ways unknown at compile time; for example when the (non-deterministic) outcome of a quantum measurement should determine which QBB would follow. Third, calling precompiled quantum routines from a QBB library will likely both reduce compilation time and improve performance, since a precompiled QBB may already be highly optimized for a specific hardware.

Thus it is imperative to reduce the latency between calls to consecutive QBBs, since the quantum system may need to maintain coherence across multiple QBBs. In the long term, quantum error correcting procedures may need to be applied between QBB calls, potentially requiring tighter synergy between the quantum processor and a controller solely dedicated to quantum error correction.

\subsection{Parameterized Quantum Instructions}
\begin{figure}[ht]
\includegraphics[width=0.9\columnwidth]{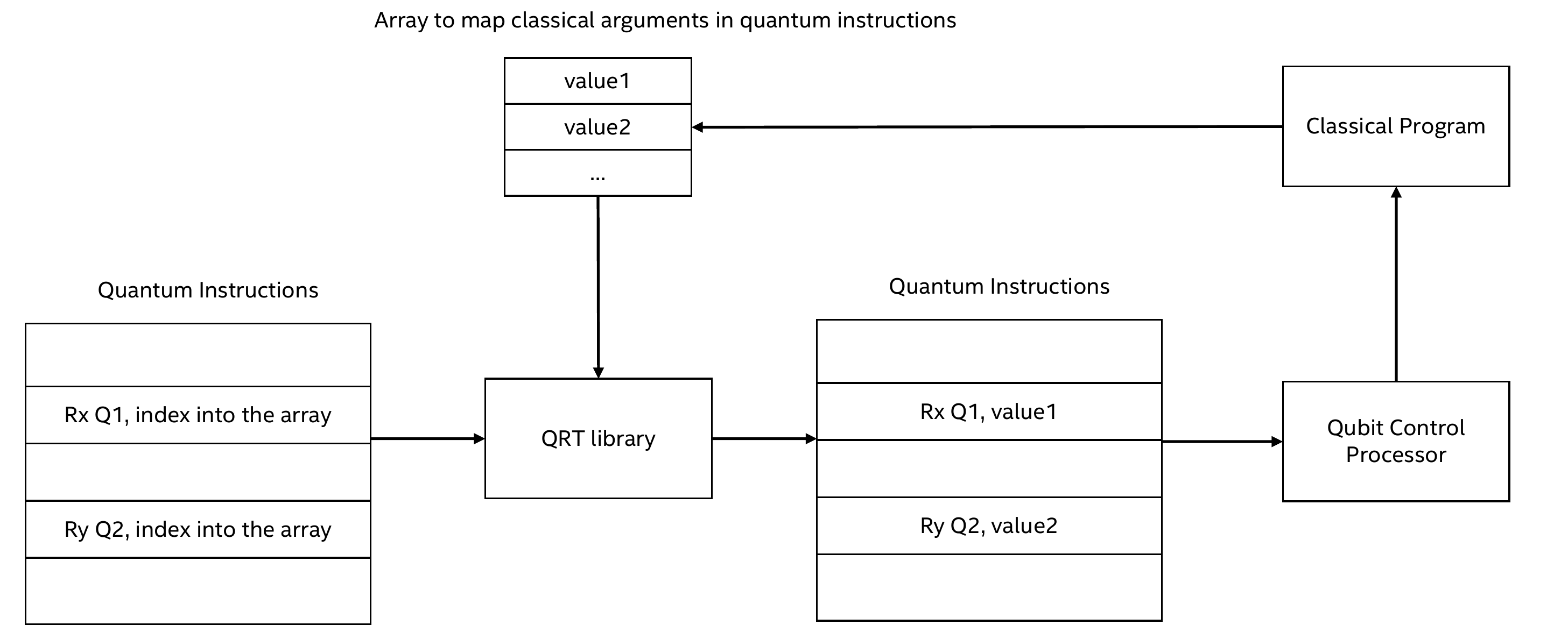}
\caption{Dynamic parameter update for parameterized instructions. The classical code running on host CPU (box in the upper left corner) computes values of the dynamic parameters during execution. The QRT library intercepts the instructions which use dynamic parameters and issues to the quantum device an instruction with the explicit value of the parameter.\label{fig:dynamic_parameters}}
\end{figure}

Our design facilitates tight integration of classical CPU processing and the quantum accelerator during runtime execution. Hence, our solution can update the instructions dynamically for parameterized circuits at runtime. This design allows users to specify both quantum and classical functions in a single program file and generates a single integrated executable binary. Our approach can leverage symbol mapping to update parameters used in a quantum circuit at runtime. \Cref{fig:dynamic_parameters} illustrates the workflow of updating quantum instructions at runtime. The host classical program updates the variables used by the quantum instructions, and the Quantum Runtime (QRT) library only updates the parameterized instructions. Unlike re-compiling the whole circuit at runtime, the QRT library only updates the variables with actual values for the parameterized instructions. With this technique, programmers only need one compilation before executing the complete variational algorithm including all iterations of the classical optimization loop. This leads to a significant performance benefit compared to traditional approaches in which each circuit instantiation requires a dedicated compilation.

\section{Implementation\label{sec:implementation}}
\begin{figure}[ht]
\includegraphics[width=0.3\columnwidth]{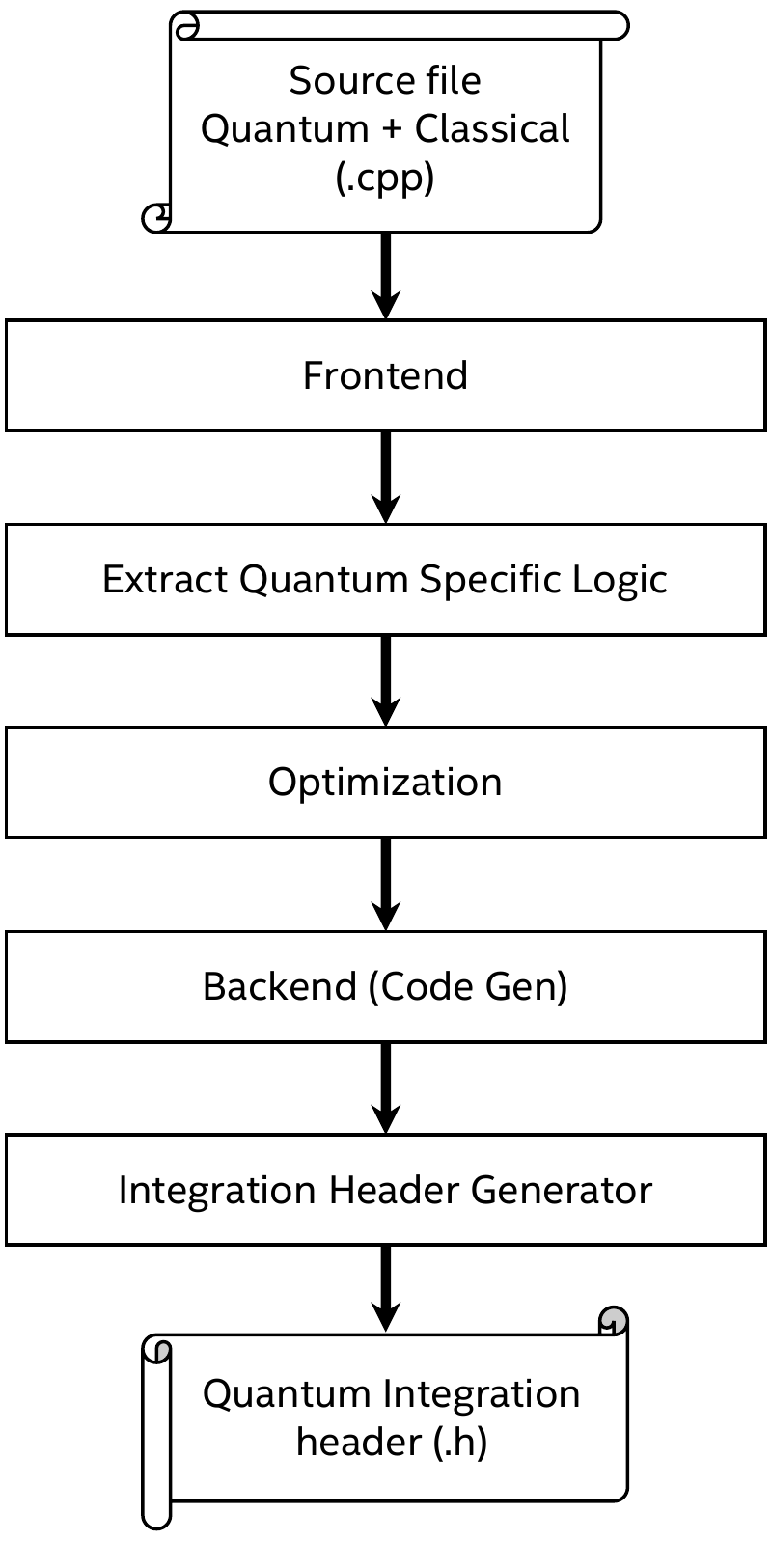}
\caption{Compilation workflow on the quantum device side to generate quantum integration header. The components of the compiler toolchain are represented with boxes, and the input-output files with scrolls. User input is distinguished with a vertical scroll and the compiler-generated file as horizontal scroll. The application source code is translated from C++ (higher-level language) to intermediate representation (IR) and then transformed by a series of optimization passes. The transformed IR is consumed by our custom quantum target backend (code generator) to emit machine code. The Integration Header Generator is an extension to LLVM tools which produces a header file to be merged back in the classical code.\label{fig:compilation_flow}}
\end{figure}

In this section we describe the key extensions to LLVM and ELF standards as part of development of our compiler toolchain, and present the quantum device compiler and its constituent elements. \Cref{fig:compilation_flow} portrays the main stages in the quantum device compilation workflow broken down into multiple steps. Finally, we show all the components of a sample hybrid quantum-classical algorithm.

\subsection{Frontend Extensions}
For our language extensions, we defined two new custom attributes - \texttt{quantum\_kernel} and \texttt{quantum\_shared\_var}. The \texttt{quantum\_kernel} attribute identifies quantum-specific functions in a program to allow intermixing of classical and quantum code in a unified source file. An array annotated with our custom attribute \texttt{quantum\_shared\_var} enables sharing of data from classical functions to quantum functions. We also use custom datatypes \texttt{qbit} and \texttt{cbit} for representing quantum operands (qubits) and classical bits, respectively. This concept of custom datatypes is adopted from the ScaffCC framework \citep{JavadiAbhari2015}. 

A standard set of quantum logic gates is made available through the \texttt{quintrinsics.h} header file that resides under the \texttt{clang/Quantum} directory in the LLVM structure. Although each quantum operation has an identifier name, a gate is defined by its matrix representation (see \Cref{code:quintrinsic}). This avoids any ambiguity in the behavior of a quantum gate, and allows easy modifications to the \emph{name} for a future standard.  This file also defines two macros \texttt{quantum\_kernel} and \texttt{quantum\_shared\_double\_array} as shorthand to apply the newly defined attributes to functions and array variables, respectively. 

\begin{listing}
\input{code_snippets/quintrinsic}
\caption{Quantum operation defined by its matrix representation. The properties of the quantum operation are captured as an annotation attribute in JSON format. The quantum operation is identified by its properties and not the name associated with the function (`X'). \label{code:quintrinsic}}
\end{listing}

The frontend is responsible for translating the source code from a higher-level language to an intermediate representation (IR). By default, an LLVM generated IR file has the \texttt{.ll} extension. At this stage, the input is a user-defined unified C++ source file with intermixed quantum-classical logic. As a result, the output is also intermixed quantum-classical IR.

\subsection{Transformation and Optimization Passes}
We utilize the LLVM Pass infrastructure to define custom transformation passes that perform IR to IR translation for the quantum program, as well as leverage  LLVM's built-in passes (\textit{e.g.} dead code elimination, instruction combining, loop unrolling, constant folding).  In addition, we have also developed our own set of embedded application programmer interfaces (APIs) for streamlining certain quantum-specific operations. This allows us to alias away some less important details for future pass-writers who are unfamiliar with the LLVM framework, while remaining flexible to implementation changes without breaking current passes. These APIs include methods for extracting quantum gate attributes, iterating through quantum-only instructions, and wrapper classes for qubit and parametric gate data. These passes and tools are used to extract, manipulate, and decompose the quantum logic. In this sub-section, we introduce a few key custom transformation passes we developed for the compilation flow.

\begin{itemize}
\item Extract quantum-specific logic: This pass takes the combined classical-quantum IR and filters it to extract quantum-only logic and its dependencies. The quantum-specific IR will be processed by the quantum device backend.
\item Optimization: This step includes a set of passes for transforming the quantum instructions to reduce resource usage on the quantum hardware, while maintaining the same quantum logic. We use typical metrics such as total number of instructions, total number of costly operations (\textit{e.g.} prioritizing multi-qubit operations over single-qubit operations), and \emph{circuit depth} (approximate time to completion on quantum hardware with parallelism considered). Though this may include common routines and methods \cite{Fagan2018, Nam2018, Nash2020}, we focus on the use of {\it product-of-Pauli-rotations} circuit synthesis \cite{schmitz2021} extended to general quantum programs \cite{schmitz2022}.
\item Gate decomposition: The standard gates are transformed into quantum operations natively supported by the quantum device as defined in the QuISA. Further optimization passes may be applied following gate decomposition into native operations. 
\item Qubit mapping: To make optimal use of available resources, the program qubits are assigned to the available physical qubits taking qubit connectivity and availability into consideration.
\item Scheduling: The sequence of quantum operations is updated considering the gate duration and timing information for the target quantum device. 
\end{itemize}

\subsection{Compiler Backend Extensions}
In the LLVM framework, the compiler backend converts the IR to the machine code of the target device. This conversion process is also referred to as \texttt{CodeGen} (\textit{i.e.} machine code generation). It involves conversion of IR to a directed acyclic graph (DAG) representation, legalization of the DAG to ensure all the datatypes are supported on the target, instruction selection to create a new DAG of machine code by means of pattern matching, register allocation to assign the physical resources on the target, and machine code emission \citep{LLVM_CodeGen}.

We implemented a custom backend in LLVM that converts the IR to machine code defined for our qubit control processor's instruction set. We defined a new class for \emph{quantum machine} which is inherited from the \texttt{TargetMachine} class \citep{LLVM_Backend}. In this phase, relocatable machine code is generated –- referred to as a quantum object file (\texttt{.qo} extension). This backend can optionally generate an assembly file for human-readable version of the machine code (\texttt{.qs} extension).

The linker is extended to add support for ELFQ and produces an executable from the quantum object file. This binary executable puts the quantum instructions in the \texttt{.qbbs\_text} section and creates a table header in \texttt{.qbbs} section.

\subsection{Quantum and Classical Code Integration}
The integration header generator tool creates a quantum integration header file that is subsequently merged back in the classical part of the code. This tool also generates metadata in the form of a mapping file that associates the functions annotated with \texttt{quantum\_kernel} attribute to their identifier in the \texttt{.qbbs} section of the ELFQ file.

LibTooling is a library provided by Clang to support source-to-source translation tools. We leverage this facility in the LLVM framework to implement a tool called \texttt{QuantumKernelReplacer}. This tool instantiates a \texttt{FrontendActionFactory} class in Clang which invokes the specified action in the Clang frontend. Using the classes \texttt{MatchFinder} and \texttt{ASTConsumer}, this tool parses the abstract syntax tree (AST) of the input source program, looks for the invocations of functions annotated with the \texttt{quantum\_kernel} attribute, and replaces them with equivalent Quantum Runtime library API calls using the mapping from the Integration Header Generator tool. Thus, the quantum compute logic is represented in a different format and merged back into the classical-only parts.

\subsection{Extensible and Linkable File for Quantum (ELFQ)}

\begin{figure}[ht]
\includegraphics[width=0.52\columnwidth]{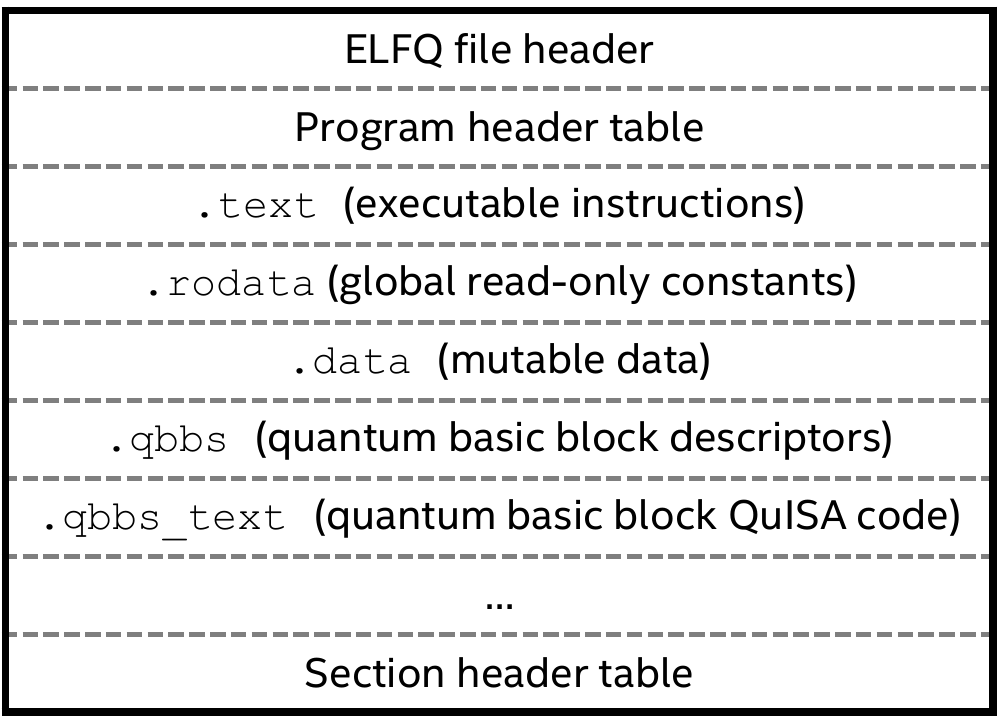}
\caption{High-level view of Extensible and Linkable File for Quantum (ELFQ) 64-bit binary file format. The two new sections defined for capturing quantum instructions are \texttt{.qbbs\_text} and \texttt{.qbbs}. \label{fig:elfq64_format}}
\end{figure}

We define extensions to the industry standard ELF format so that a program executable binary generated by our compiler toolchain carries with it both the classical and the quantum program code in 64-bit binary form. \Cref{fig:elfq64_format} gives a high-level view of an ELFQ file. The quantum kernels are compiled into binary instructions and placed in the quantum basic blocks section (\texttt{.qbbs\_text}). A table header in the \texttt{.qbbs} section is generated that holds an identifier for each quantum kernel, its size, alignment, and offset. A key part of our quantum runtime execution model is the arrangement, selection, and issuance of quantum basic blocks to the quantum accelerator during the standard flow of the classical program. It is achieved by the operating system program loader resolving the location of the quantum basic block table header.

\subsection{Quantum Runtime}

\begin{figure}[ht]
\includegraphics[width=0.65\columnwidth]{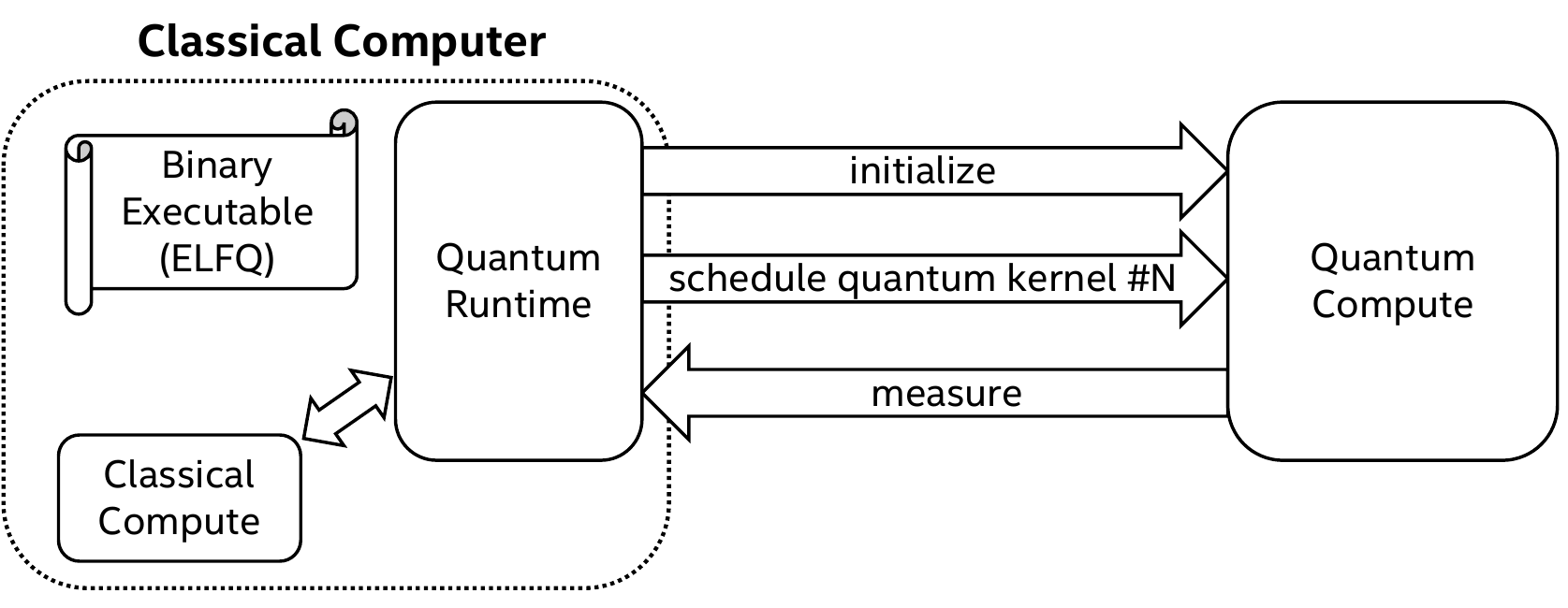}
\caption{The Quantum Runtime (QRT) manages the interaction between the classical computer and quantum device.\label{fig:QRT}}
\end{figure}

The Quantum Runtime (QRT) library provides an API for initialization of the underlying quantum target, scheduling of quantum functions synchronously or asynchronously to the quantum target device, and retrieval of results from quantum measurement operations. The QRT is also responsible for resolving parameters that are unknown at compile time and computed at runtime. This allows execution of variational algorithms without having to recompile the application. However, the ansatz and the measurement scheme must be defined at compile time. The interaction between the classical host computer and the quantum device as managed by the QRT is depicted in \Cref{fig:QRT}.

\subsection{Sample Application Program}

\begin{listing}
\input{code_snippets/sample_hybrid_program}
\caption{A sample program demonstrating the usage of dynamic quantum parameters for variational algorithms\label{code:sample_program}}
\end{listing}

\Cref{code:sample_program} shows a sample test program which uses these newly defined C++ extensions. To write a new application to use our toolchain, implement the following steps: 
\begin{enumerate}
\item Include the \texttt{quintrinsics.h} header file to use any of the pre-defined standard gates: 
\begin{align*}
\left\{ X, Y, Z, H, S, S^\dagger, T, T^\dagger, \mathsf{C}Z, \mathsf{CNOT}, \mathsf{SWAP}, \mathsf{CCNOT}, \mathsf{R}_X{(\theta)}, \mathsf{R}_Y{(\theta)}, \mathsf{R}_Z{(\theta)} \right\}
\end{align*}
\item Define qubits as global variables – single or multiple, individual or arrays or any combination. Similarly, define classical bits for capturing measurement results. The number of available qubits depends on the target quantum device or simulator. 
\item Define an array of desired size, annotated with the attribute \texttt{quantum\_shared\_var}, to hold any dynamic parameters that will be modified during runtime. 
\item Annotate the functions which encapsulate quantum operations with the \texttt{quantum\_kernel} attribute. 
\item Define compute-intensive logic, including the functionality to compute values for dynamic parameters, and classical code in conventional C++ functions. 
\end{enumerate}

\section{Example Workload\label{sec:workload}}

\subsection{Workload Overview}

The example workload considered in this manuscript is a hybrid quantum-classical variational algorithm for the generation of thermofield double (TFD) states \citep{Wu2019, Premaratne2020, Sagastizabal2021}, executed on the Intel quantum compiler toolchain, quantum runtime library, and Intel Quantum Simulator. To demonstrate the capabilities of this full-stack in simulation, we generated TFD states of size $\Nq \in \{4,6,8,10,12,14,16,18,20\}$. We validated the results using an independent algorithm implementation in MATLAB for system sizes up to $\Nq=12$. 

\subsection{Algorithm Details}

A thermal state is a particular form of a mixture of pure states \citep{Nielsen2010}. Generating thermal states is especially important for applications such as modeling condensed matter systems. Since thermal states are a result of non-unitary evolution of a quantum system, it is necessary to explore alternative methods to simulate thermal states in a quantum computer. A TFD state is a specific purification of a thermal state (or Gibbs state) \citep{Wu2019}. To prepare a thermal state of an $L$-qubit system, two subsystems (each containing $L$ qubits) are required by the TFD generation algorithm \citep{Wu2019}. Entanglement between the two subsystems is generated via multi-qubit operations in such a way that, when one of the subsystems is traced out, the remaining subsystem is prepared in the desired mixed state. Here, we focus on preparation of TFD states for the transverse-field Ising model (TFIM) in a one-dimensional ring described by the following Hamiltonian,
\begin{align}
H_\textrm{TFIM} = \sum_{i=1}^{L} Z_i Z_{i+1} + g \sum_{i=1}^L X_i \equiv H_{ZZ} + g H_X 
\label{eq:H_TFIM}
\end{align}
where $Z_i$ and $X_i$ are Pauli operators acting on the $i^\textrm{th}$ qubit, and $g$ is the transverse field strength \citep{Premaratne2020}.

\begin{figure}[ht]
\includegraphics[width=0.7\columnwidth]{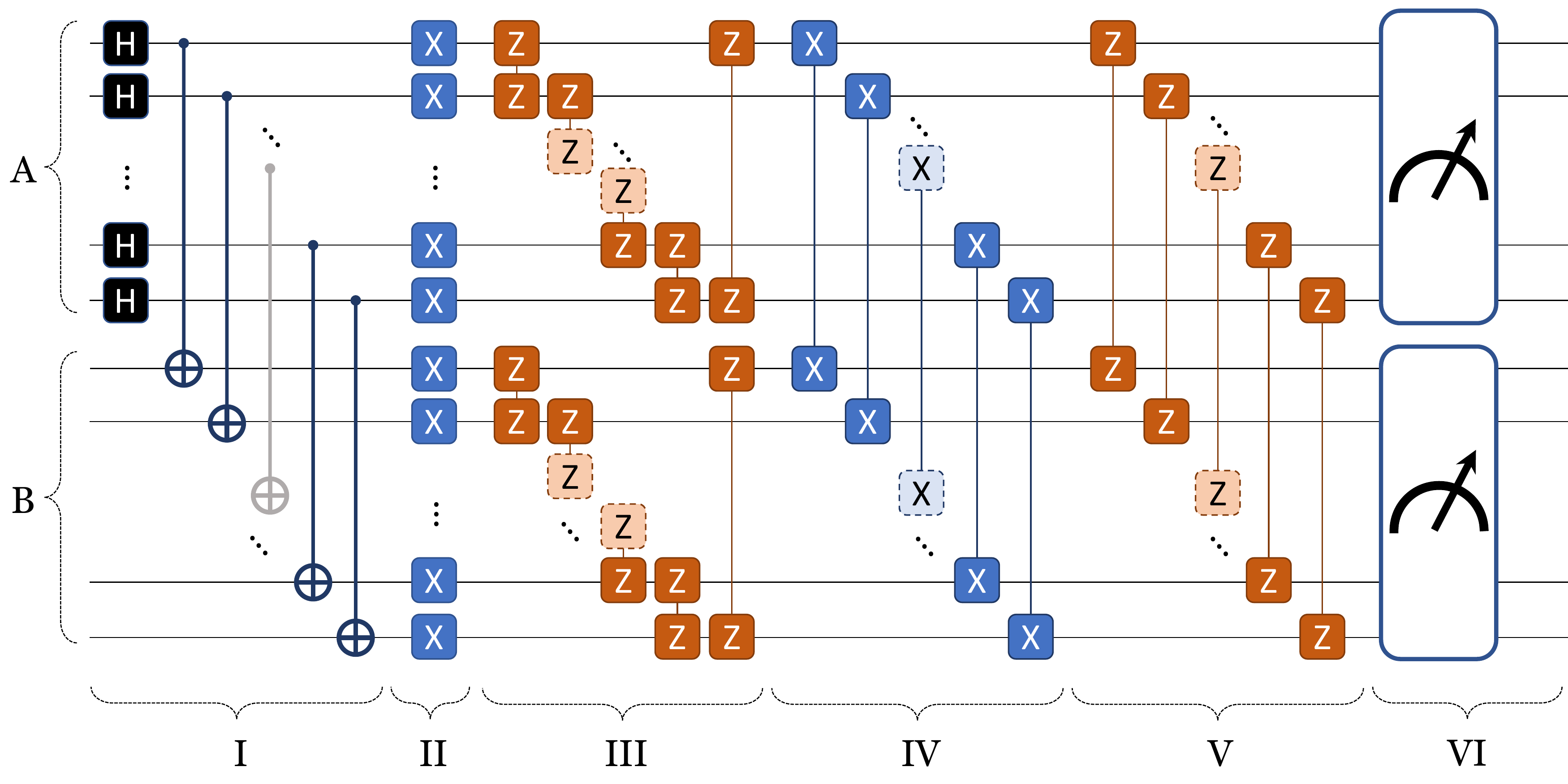}
\caption{The quantum circuit for single-step TFD state generation. Stages: (I) preparing infinite temperature TFD state, (II) Intra-system $R_X$ operation with angle $\gamma_1$, (III) Intra-system $ZZ$ operation with angle $\gamma_2$, (IV) Inter-system $XX$ operation with angle $\alpha_1$,  (V) Inter-system $ZZ$ operation with angle $\alpha_2$ (VI) multi-qubit measurements to verify TFD generation. A and B represent the two subsystems, each containing $L$ qubits. \label{fig:ansatz}}
\end{figure}

We limit ourselves to the case of $g = 1$ due to the existence of a critical point in the behavior of the system \citep{Bonfim2019, Pfeuty1970}. The process begins by preparing pairwise Bell states between corresponding qubits in the two subsystems (A and B) each containing $L$ qubits. This realizes the infinite temperature TFD states. Then, intra-system operators corresponding to the exponentiation of \cref{eq:H_TFIM} via the Trotter method \citep{Yung2014} are applied within each of the subsystems, followed by mixing/entangling operations between the subsystems. Each of the intra-system and inter-system operations have associated variational angles that can be classically optimized for generating a good approximation of the desired TFD state. The full quantum circuit for a variable number of qubits is shown in \cref{fig:ansatz}.

The degree of mixture/entanglement is dependent on the desired temperature, and effectively encoded within the calculation of the cost function. The results discussed below are obtained based on single-step TFD generation (four variational angles) \citep{Premaratne2020}, but for higher fidelities with higher qubit numbers it is desirable to have more steps and consequently more variational angles to be optimized \citep{Wu2019}. In this work, the four variational angles are labeled {$\gamma_1$, $\gamma_2$, $\alpha_1$, $\alpha_2$} as described in \cref{fig:ansatz}. 

The ideal cost function for generating thermal states is the Gibbs free energy \citep{Wu2019}. However, evaluating the Gibbs free energy is a costly procedure on a quantum processor, and it is desirable to have approximate cost functions that can be efficiently evaluated \citep{Premaratne2020}. In this work we utilize such an approximate cost function to evaluate the effectiveness of the generated states:
\begin{align}
\begin{split}
\mathcal{C}(\beta) &= \expval{\sum_{i=1}^L {X_i^A}} + \expval{\sum_{i=1}^L {X_i^B}} + \expval{\sum_{i=1}^L {Z_i^A Z_{i+1}^A}} + \expval{\sum_{i=1}^L {Z_i^B Z_{i+1}^B}} \\
&\hspace{4em} - \beta^{-1} \left( \expval{\sum_{i=1}^L {X_i^A X_i^B}} + \expval{\sum_{i=1}^L {Z_i^A Z_i^B}} \right) \\
&= \expval{X_A} + \expval{X_B} + \expval{{ZZ}_A} + \expval{{ZZ}_B} - \beta^{-1} \left( \expval{{XX}_{AB}} + \expval{{ZZ}_{AB}} \right)
\end{split}
\end{align}
where the single subscript operators are acting within the corresponding subsystem and the double subscript operators represent the operators acting between the subsystems A and B \citep{Premaratne2020}. Here $\beta$ is the inverse temperature, and the expectation values are evaluated with respect to the generated state. The minimum possible value of our cost function $\mathcal{C}(\beta)$ depends on both the system size and the targeted temperature. The lowest possible $\mathcal{C}(\beta)$ reached via classical optimization identifies the final approximation to the TFD state. To quantify its accuracy, we compute the fidelity with the desired thermal state.

\subsection{Quantum Programming Considerations}

Using the Intel Quantum Simulator, we are able to access the final state vector at the end of quantum system evolution. If we were allowed to use the full knowledge of the state vector, then computing $\mathcal{C}(\beta)$ would require only a single simulation (as opposed to many experimental repetitions required to accumulate measurement statistics). However, for this demonstration we use the concept of the Probability Register, which stores the probabilities for each unique basis state combination (as opposed to its complex probability amplitude). For example, in a three-qubit system, a single circuit evaluation using IQS will give us access to the probabilities for each of the following states:
$\ket{000}$, $\ket{001}$, $\ket{010}$, $\ket{011}$, $\ket{100}$, $\ket{101}$, $\ket{110}$, $\ket{111}$.

With this information available after each circuit evaluation, we will require two runs of the same circuit for each optimization iteration. In the first run, the measurements will be performed in the conventional $Z$-basis allowing us to calculate the $Z$ related operators in $\mathcal{C}(\beta)$. In the second run, Hadamard operations will be applied to all qubits prior to measurements resulting in $X$-basis measurements that will allow us to calculate the remaining X related expectation values in $\mathcal{C}(\beta)$.
The optimization is performed using the popular \texttt{dlib} library \citep{dlib}. This library contains several different optimization routines, and in this work we found that the \texttt{find\_min\_bobyqa} function performs well. See the \Cref{app:code_snippets} for code snippets of the example.

\subsection{Results}
The accuracy of TFD state generation is evaluated by calculating the fidelity of the generated states with respect to the ideal thermal state at each temperature. First, the pure state on $\Nq = 2L$ qubits is reduced to a density matrix, corresponding to an $L$ qubit mixed state, by tracing over subsystem B to yield $\rho_\textrm{sim}$. Then, the thermal state accuracy is calculated to be the fidelity between the ideal ($\rho_\textrm{ideal}$) and generated density matrices ($\rho_\textrm{sim}$) as 
\begin{align}
\mathcal{F} = \left[ \mathrm{Tr} \sqrt{ \rho_\textrm{ideal}^{0.5} \rho_\textrm{sim} \rho_\textrm{ideal}^{0.5} } \right]^2
\end{align}

\begin{figure}[t]
\includegraphics[width=0.65\columnwidth]{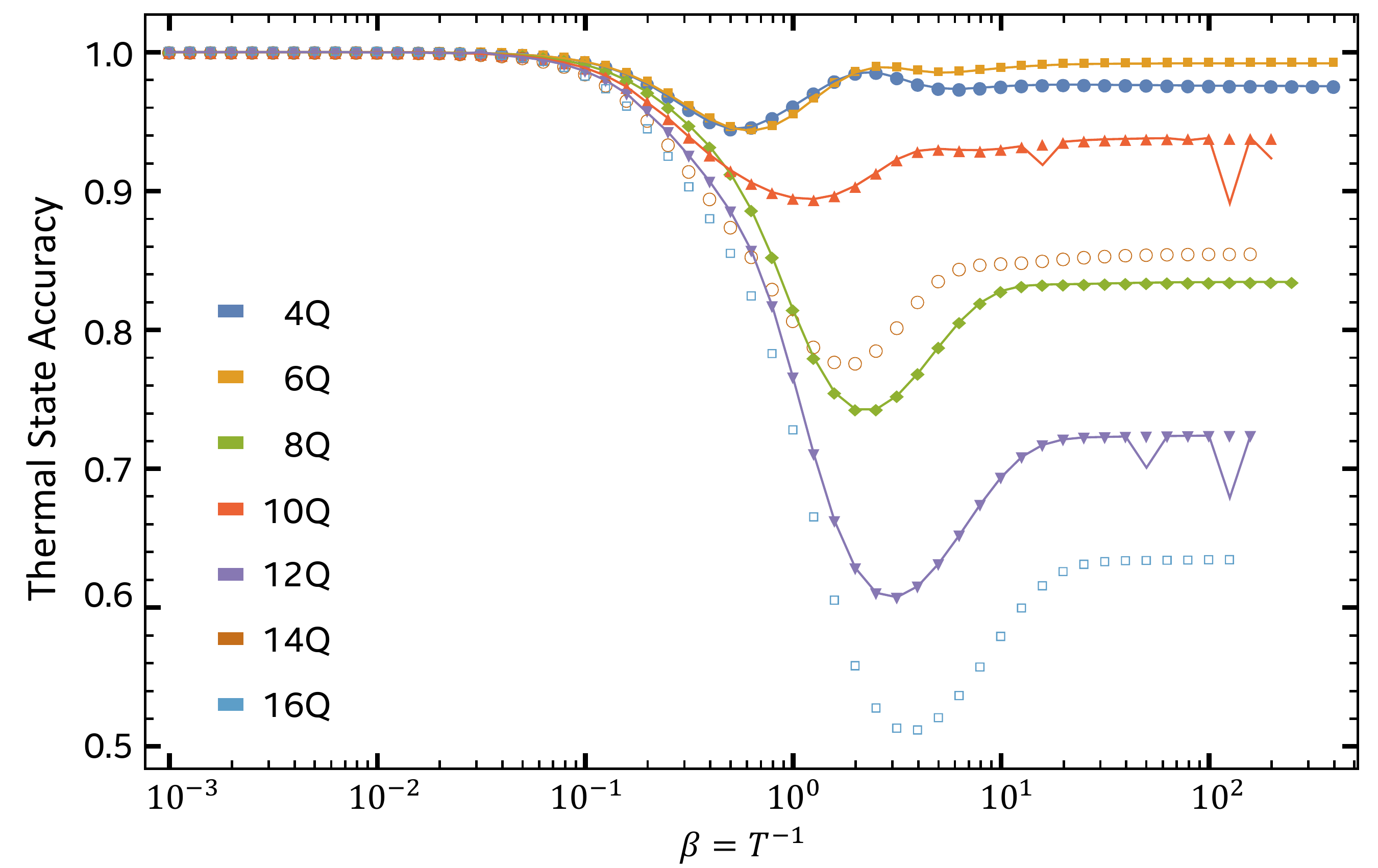}
\caption{Comparison of the thermal state accuracy for variational optimization performed in two different ways. The markers show data from using the Intel compiler toolchain plus Intel Quantum Simulator. The solid lines show data from using MATLAB. Note that the number in parentheses represents $\Nq (=2L)$.\label{fig:sdk_vs_matlab}}
\end{figure}

We observe excellent agreement up to $\Nq = 12$ between the optimizations when comparing the results obtained using this full-stack in simulation and the independent MATLAB scripts utilizing the \texttt{fminsearch} function. Beyond this system size, the MATLAB implementation slowed down considerably. The non-monotonic trend in the lowering of fidelities for low temperatures with increasing system size continued as expected for $\Nq > 12$ (see \cref{fig:sdk_vs_matlab}). The calculation of thermal state accuracy was performed using a MATLAB script, and for $\Nq > 16$ this analysis was found to be quite time consuming. More sophisticated analysis and verification techniques are expected to enable studies for larger system sizes. There are several general trends that can be observed in the results, but a thorough analysis and interpretation of these aspects is beyond the scope of this work.



\section{Conclusion\label{sec:conclusion}}
	
A full-stack framework to compile and execute quantum programs is key to harnessing the power of quantum computation. In this paper, we introduced an Intel-developed C++ compiler toolchain with quantum extensions, based on the LLVM framework. Our framework is specifically designed to support efficient compilation and execution of hybrid quantum-classical algorithms. Some of the strongest candidates to achieve quantum advantage are expressed in the form of hybrid programs like VQE and QAOA, and even tighter interaction between host computers and quantum accelerators is expected to enable end-to-end solutions for real applications. For this important class of applications our approach requires only one compilation, a key advantage enabled by supporting dynamically-determined parameters. This significantly reduces the execution latency of hybrid workloads.

To showcase the framework's functionalities and discuss the frontend's extension in a concrete setting,  we demonstrate the execution of a hybrid algorithm through this full-stack in simulation, using the high-performance Intel Quantum Simulator in place of a quantum accelerator. The modular structure of the proposed full-stack facilitates its extension to target different qubit backends in the near future. These may include simulators reproducing the behavior of quantum hardware at a more physical level. For example, a quantum dot simulator aware of the effects of multiplexed microwave drive pulses on electron spins would contribute to the design of practical applications tailored for the quantum dot qubit technology. Eventually, this toolchain will be used to drive physical control electronics to execute quantum kernels on physical qubit devices.

\begin{acknowledgments}
The authors thank Nicolas P.D. Sawaya for insightful discussions and comments on the manuscript. The authors specially acknowledge James P. Held for his guidance, wisdom, and support.
\end{acknowledgments}

\bibliographystyle{apsrev4-2}
\bibliography{all_refs}

\newpage

\appendix*

\section{Example Workload Code Snippets\label{app:code_snippets}}

\subsection{Initialization}
 
In the initialization section we define the number of qubits and the other essential parameters to construct a particular-sized quantum circuit. This section also defines the data structures which are used to retrieve the measurement outcomes as well as those to perform an efficient classical optimization. See \Cref{code:init}.

\begin{listing}
\input{code_snippets/init}
\caption{Initialization of parameters, and definition of necessary data structures\label{code:init}}
\end{listing}

\subsection{Quantum Kernel for Variational Terms}
 
The \texttt{TFD\_terms} quantum kernel is the heart of the quantum circuit, and is where the operations related to the variational terms are defined. It is generally written such that it can accommodate any number of qubits in the system, as well as several TFD steps. See \Cref{code:tfdq4_terms}.

\begin{listing}
\input{code_snippets/tfdq4_terms}
\caption{Quantum kernel containing the core of the ansatz of TFD generation\label{code:tfdq4_terms}}
\end{listing}

\subsection{Supporting Quantum Kernels}
 
The supporting quantum kernels are used to construct the full quantum kernel that will be invoked within the main function. These are used to organize simpler groups of operations on many qubits. See \Cref{code:aux_qkernels}.

\begin{listing}
\input{code_snippets/aux_qkernels}
\caption{Individual quantum kernels that are dedicated for well-defined steps when running an experiment\label{code:aux_qkernels}}
\end{listing}

\subsection{Composite Quantum Kernels}

The composite quantum kernels are composed of the specialized quantum kernels defined earlier. The \texttt{tfd\_Z} and \texttt{tfd\_X} quantum kernels will be directly called within the main function during execution. See \Cref{code:main_qkernels}.

\begin{listing}
\input{code_snippets/main_qkernels}
\caption{Final quantum kernels that will be run during optimization\label{code:main_qkernels}}
\end{listing}

\subsection{Cost Function Terms}
 
Here we define the cost function that will be minimized during optimization. In this example, the weighted sum of the Probability Register terms is calculated independently for defining the components of the total cost function. Hence, the cost function shown in \Cref{code:cost_function_terms} is for the special case of $\Nq = 6$.

\begin{listing}
\input{code_snippets/cost_function_terms}
\caption{Functions for calculating the cost function. Note that these expressions for the cost function terms are applicable only for the special case of $\Nq = 6$. The parameters $\left\{ \mathtt{P}, \mathtt{P\_Z}, \mathtt{P\_X} \right\}$ represent Probability Registers that are used as inputs for the cost function evaluation. \label{code:cost_function_terms}}
\end{listing}

\subsection{Main}
 
The main function expresses the full optimization routine over a range of temperatures to generate TFD states. The objective function is coded in the form of a lambda function, within which the two different quantum kernels (\texttt{tfd\_Z} and \texttt{tfd\_X}) are run and the results are used to calculate the total cost. Then, the \texttt{find\_min\_bobyqa} function from the \texttt{dlib} library is used to perform the minimization on the objective function. See \Cref{code:main}.

\begin{listing}
\input{code_snippets/main}
\caption{The \texttt{main} function\label{code:main}}
\end{listing}

\end{document}

%% file: code_snippets/quintrinsic.tex
\begin{minted}[bgcolor=bg]{c++} 
/* Pauli-X gate */
void X(qbit q) __attribute__((annotate("{ \
    \"matrix_real\" : [0, 1, 1, 0], \
    \"matrix_imag\" : [0, 0, 0, 0], \
    \"matrix_order\" : \"rm\", \
    \"is_hermitian\" : true, \
    \"is_unitary\" : true, \
    \"is_mutable\" : true, \
    \"qubit_list\" : [0], \
    \"parametric_list\" : [], \
    \"control_qubit_list\" : [], \
    \"local_basis_list\" : [1], \
    \"identifier\" : 2 \
  }"))) {
  __quantum_qubit(q);
}
\end{minted}

%% file: code_snippets/sample_hybrid_program.tex
\begin{minted}[bgcolor=bg]{c++} 
#include <clang/Quantum/quintrinsics.h>
#include <iostream>
#include <math.h>
#include <time.h>

const int N = 3;
qbit q[N];
cbit c[N];

/* Array to hold dynamic parameters for quantum algorithm */
quantum_shared_double_array MyVariableParams[3];

quantum_kernel void prepAll() { for (int i = 0; i < N; i++) { PrepZ(q[i]); }}

quantum_kernel void measAll() { for (int i = 0; i < N; i++) { MeasZ(q[i], c[i]); }}

quantum_kernel void qfoo() {
  RX(q[0], MyVariableParams[0]);
  RY(q[1], MyVariableParams[1]);
  RZ(q[2], MyVariableParams[2]);
}

int main() {
  srand(time(0));  
  prepAll();
  /* Dynamically update parameters multiple times */
  for (int i = 0; i < N; i++) {
    MyVariableParams[0] = static_cast<double>(rand()) / (RAND_MAX / M_PI);
    MyVariableParams[1] = std::pow(MyVariableParams[0], 1.1);
    MyVariableParams[2] = std::sqrt(MyVariableParams[0]);
    qfoo();
  }
  measAll();
  for (int i = 0; i < N; i++) { std::cout << "Qubit#" << i << " : " << (bool)c[i] << "\n"; }  
  return 0;
}

/* 
 * Sample console output *
Qubit#0 : 0
Qubit#1 : 1
Qubit#2 : 0
*/
\end{minted}

%% file: code_snippets/init.tex
\begin{minted}[bgcolor=bg]{c++} 
#include <clang/Quantum/quintrinsics.h>

#include <vector>
#include <math.h>
#include <dlib/optimization.h>

/* Define the number of qubits needed for compilation */
const int N = 6;              // Total number of qubits
const int N_ss = 2;           // Number of subsystems
const int N_sub = (int)(N/N_ss);  // Number of qubits in subsystem

const int N_var_angles = 4;
qbit QReg[N];
cbit CReg[N];

/* QRT Provided data structure to get state probabilities */
extern std::vector<double> ProbabilityRegister;

/* Array to hold dynamic parameters for quantum algorithm */
quantum_shared_double_array QVarParams[N_var_angles];

/* Defining a column vector for storing the input to objective functions to be minimized */
typedef dlib::matrix<double, N_var_angles, 1> column_vector;
\end{minted}

%% file: code_snippets/tfdq4_terms.tex
\begin{minted}[bgcolor=bg]{c++} 
quantum_kernel void TFD_terms () {

  /* This is to shift the variational parameter in case of more steps */
  int shift = 0;

  int idxSysX = 0;
  int idxSysZ = 0;
  int idxMixX = 0;
  int idxMixZ = 0;
  
  const double PiOver2 = 1.57079632679;

  /* Single qubit variational terms */
  for (idxSysX = 0; idxSysX < N; idxSysX++)
    RX(QReg[idxSysX], QuantumVariableParams[0 + shift]);

  /* Two-qubit intra-system variational terms (adjacent) */
  for (int idxGrandZ = 0; idxGrandZ < N_sub - 1; idxGrandZ++) {
    for (idxSysZ = 0; idxSysZ < N_ss; idxSysZ++)
      CNOT(QReg[idxGrandZ + N_sub * idxSysZ + 1], QReg[idxGrandZ + N_sub * idxSysZ]);
    for (idxSysZ = 0; idxSysZ < N_ss; idxSysZ++)
      RZ(QReg[idxGrandZ + N_sub * idxSysZ], QuantumVariableParams[1 + shift]);
    for (idxSysZ = 0; idxSysZ < N_ss; idxSysZ++)
      CNOT(QReg[idxGrandZ + N_sub * idxSysZ + 1], QReg[idxGrandZ + N_sub * idxSysZ]);
  }
  /* Two-qubit intra-system variational terms (boundary term) */
  for (idxSysZ = 0; idxSysZ < N_ss; idxSysZ++)
    CNOT(QReg[N_sub * idxSysZ], QReg[N_sub * idxSysZ + (N_sub - 1)]);
  for (idxSysZ = 0; idxSysZ < N_ss; idxSysZ++)
    RZ(QReg[N_sub * idxSysZ + (N_sub - 1)], QuantumVariableParams[1 + shift]);
  for (idxSysZ = 0; idxSysZ < N_ss; idxSysZ++)
    CNOT(QReg[N_sub * idxSysZ], QReg[N_sub * idxSysZ + (N_sub - 1)]);

  /* Two-qubit inter-system XX variational terms */
  for (idxMixX = 0; idxMixX < N_sub; idxMixX++) {
    RY(QReg[idxMixX + N_sub], -PiOver2);
    RY(QReg[idxMixX], -PiOver2);
  }
  for (idxMixX = 0; idxMixX < N_sub; idxMixX++)
    CNOT(QReg[idxMixX + N_sub], QReg[idxMixX]);
  for (idxMixX = 0; idxMixX < N_sub; idxMixX++)
    RZ(QReg[idxMixX], QuantumVariableParams[2 + shift]);
  for (idxMixX = 0; idxMixX < N_sub; idxMixX++)
    CNOT(QReg[idxMixX + N_sub], QReg[idxMixX]);
  for (idxMixX = 0; idxMixX < N_sub; idxMixX++) {
    RY(QReg[idxMixX + N_sub], PiOver2);
    RY(QReg[idxMixX], PiOver2);
  }

  /* Two-qubit inter-system ZZ variational terms */
  for (idxMixZ = 0; idxMixZ < N_sub; idxMixZ++)
    CNOT(QReg[idxMixZ], QReg[idxMixZ + N_sub]);
  for (idxMixZ = 0; idxMixZ < N_sub; idxMixZ++)
    RZ(QReg[idxMixZ + N_sub], QuantumVariableParams[3 + shift]);
  for (idxMixZ = 0; idxMixZ < N_sub; idxMixZ++)
    CNOT(QReg[idxMixZ], QReg[idxMixZ + N_sub]);

}
\end{minted}

%% file: code_snippets/aux_qkernels.tex
\begin{minted}[bgcolor=bg]{c++} 
quantum_kernel void PrepZAll () {
  /* Initialization of the qubits */
  for (int idx = 0; idx < N; idx++)
    PrepZ(QReg[idx]);
}

quantum_kernel void BellPrep () {
 /* Preparation of Bell pairs (T -> Infinity) */
  for (int idx = 0; idx < N_sub; idx++)
    RY(QReg[idx], PiOver2);
  for (int idx = 0; idx < N_sub; idx++)
    CNOT(QReg[idx], QReg[idx + N_sub]);
}

quantum_kernel void MeasZAll () {
  /* Measurements of all the qubits */
  for (int idx = 0; idx < N; idx++)
    MeasZ(QReg[idx], CReg[idx]);
}

quantum_kernel void XmaptoZ () {
  /* Mapping from X basis to Z basis */
  for (int idx = 0; idx < N; idx++)
    H(QReg[idx]);
}
\end{minted}

%% file: code_snippets/main_qkernels.tex
\begin{minted}[bgcolor=bg]{c++} 
/* First kind of experiment that needs to be run per iteration of optimization.  */
quantum_kernel void tfd_Z() {
  PrepZAll();
  BellPrep();
  TFD4q_terms();
  MeasZAll();
}

/*
 Second kind of experiment that needs to be run per iteration of optimization. Almost identical to the 
 previous quantum kernel tfdQ4_Z. Four Hadamard operations are inserted just prior to measurement of the 
 qubits. This maps from Z-basis to X-basis, so that X-related observables can be calculated.
*/
quantum_kernel void tfd_X() {
  PrepZAll();
  BellPrep();
  TFD4q_terms();
  XmaptoZ();
  MeasZAll();
}
\end{minted}

%% file: code_snippets/cost_function_terms.tex
\begin{minted}[bgcolor=bg]{c++} 
/* 
 The three expectation values of the observables have been written as functions below. 
 The relationships between expectation values and state probabilities was derived independently 
 with simple matrix calculations of Pauli matrix products.
*/
double ZZApZZB_expectation (std::vector<double> P) {
  double sum = 3*P[0] + P[1] + P[2] + P[3] + P[4] + P[5] + P[6] + 3*P[7] + P[8] - P[9] - P[10] \
  	- P[11] - P[12] - P[13] - P[14] + P[15] + P[16] - P[17] - P[18] - P[19] - P[20] \
  	- P[21] - P[22] + P[23] + P[24] - P[25] - P[26] - P[27] - P[28] - P[29] - P[30] \
  	+ P[31] + P[32] - P[33] - P[34] - P[35] - P[36] - P[37] - P[38] + P[39] + P[40] \
  	- P[41] - P[42] - P[43] - P[44] - P[45] - P[46] + P[47] + P[48] - P[49] - P[50] \
  	- P[51] - P[52] - P[53] - P[54] + P[55] + 3*P[56] + P[57] + P[58] + P[59] + P[60] \
  	+ P[61] + P[62] + 3*P[63];
  return 2*sum;
}

double ZZAB_expectation (std::vector<double> P) {
  return 3*P[0] + P[1] + P[2] - P[3] + P[4] - P[5] - P[6] - 3*P[7] + P[8] + 3*P[9] - P[10] \
  	+ P[11] -  P[12] + P[13] - 3*P[14] - P[15] + P[16] - P[17] + 3*P[18] + P[19] - P[20] \
  	- 3*P[21] + P[22] - P[23] -  P[24] + P[25] + P[26] + 3*P[27] - 3*P[28] - P[29] - P[30] \
  	+ P[31] + P[32] - P[33] - P[34] - 3*P[35] + 3*P[36] + P[37] + P[38] - P[39] - P[40] \
  	+ P[41] - 3*P[42] - P[43] + P[44] + 3*P[45] - P[46] + P[47] - P[48] - 3*P[49] + P[50] \
  	- P[51] + P[52] - P[53] + 3*P[54] + P[55] - 3*P[56] - P[57] - P[58] + P[59] - P[60] \
  	+ P[61] + P[62] + 3*P[63];
}

double ZApZB_expectation (std::vector<double> P) {
  double sum = 3*P[0] + 2*P[1] + 2*P[2] + P[3] + 2*P[4] + P[5] + P[6] + 2*P[8] + P[9] + P[10] \
  	+ P[12] - P[15] + 2*P[16] + P[17] + P[18] + P[20] - P[23] + P[24] - P[27] - P[29] - P[30] \
  	- 2*P[31] + 2*P[32] + P[33] + P[34] + P[36] - P[39] + P[40] - P[43] - P[45] - P[46] \
  	- 2*P[47] + P[48] - P[51] - P[53] - P[54] - 2*P[55] - P[57] - P[58] - 2*P[59] - P[60] \
  	- 2*P[61] - 2*P[62] - 3*P[63];
  return 2*sum;      
}

/*
 The total cost function is constructed here using the expectation values from before. 
 There is freedom in the numeric coefficient of the ZZ term and exponent for beta used.
*/
double total_cost (double inv_temp, std::vector<double> P_Z, std::vector<double> P_X) {
  double energy_X = ZApZB_expectation(P_X); 
  double energy_ZZ = ZZApZZB_expectation(P_Z);
  double entropy = ZZAB_expectation(P_X) + ZZAB_expectation(P_Z);
  return energy_X + 1.00 * energy_ZZ - pow(inv_temp, -1.00) * entropy;
}
\end{minted}

%% file: code_snippets/main.tex
\begin{minted}[bgcolor=bg]{c++} 
int main() {
  /*
   initial starting point. Defining it here means I will reuse the best result from
   from previous temperature when starting the next temperature run
  */
  column_vector starting_point = {0, 0, 0, 0};

  /* Looping over the idx for range of inverse temperatures */
  for (int beta_idx = -30; beta_idx < 31; beta_idx+=1) {
    /* calculating the actual inverse temperature that is used during calculations */
    double beta = pow(10, (double) beta_idx/10);

    /* Constructing a lambda function to be used for a single optimization iteration */
    auto ansatz_run_lambda = [&](const column_vector& m) {
        /* loading the new variational angles into the special global array for hybrid compilation */
        for (int j = 0; j < N_var_angles; j++)
          QVarParams[j] = m(j);

        /* two local variables to store the results from the two different experiments per iteration */
        std::vector<double> PRZ;
        std::vector<double> PRX;
        
        /* performing the Z-focused experiment, and storing the data in PRZ */
        tfd_Z();
        for (auto j = 0; j < ProbabilityRegister.size(); j++)
          PRZ.push_back(ProbabilityRegister[j]);
        
        /* performing the X-focused experiment, and storing the data in PRZ */
        tfd_X();
        for (auto j = 0; j < ProbabilityRegister.size(); j++)
          PRX.push_back(ProbabilityRegister[j]);
   
        return total_cost(beta, PRZ, PRX);
    };

    /* running the full optimization for a given temperature */
    auto result = dlib::find_min_bobyqa(
      ansatz_run_lambda, starting_point, 
      2 * N_var_angles + 1, // number of interpolation points
      dlib::uniform_matrix<double>(N_var_angles, 1, -7.0), // lower bound constraint
      dlib::uniform_matrix<double>(N_var_angles, 1, 7.0), // upper bound constraint
      1.5, // initial trust region radius
      1e-5, // stopping trust region radius
      10000 // max number of objective function evaluations
    );
  }

  return 0;
}
\end{minted}